\documentclass[useAMS,usenatbib]{mnras}
\usepackage{amsfonts}
\usepackage{amsmath}
\usepackage{amssymb}
\usepackage{graphicx}
\usepackage{float}
\usepackage{cleveref}
\usepackage{color}
\usepackage{caption}

\title[Two-fluid dusty shocks]{Two-fluid dusty shocks: simple benchmarking problems and applications to protoplanetary discs}

\author[Andrew Lehmann and Mark Wardle]{Andrew Lehmann\thanks{E-mail:
andrew.lehmann@mq.edu.au} and Mark Wardle\\
Department of Physics and Astronomy, and Research Centre for Astronomy, Astrophysics \& Astrophotonics,\\
Macquarie University, Sydney, NSW 2109, Australia}

\newcommand{\dustyshocks}{}

\newcommand{\imformat}{eps}
\newcommand{\imformattwo}{eps}

\begin{document}%

\pagerange{\pageref{firstpage}--\pageref{lastpage}} \pubyear{2015}

\maketitle

\begin{abstract}
The key role that dust plays in the interstellar medium has motivated the development of numerical codes designed to study the coupled evolution of dust and gas in systems such as turbulent molecular clouds and protoplanetary discs. Drift between dust and gas has proven to be important as well as numerically challenging. We provide simple benchmarking problems for dusty gas codes by numerically solving the two-fluid dust-gas equations for steady, plane-parallel shock waves. The two distinct shock solutions to these equations allow a numerical code to test different forms of drag between the two fluids, the strength of that drag and the dust to gas ratio. We also provide an astrophysical application of J-type dust-gas shocks to studying the structure of accretion shocks onto protoplanetary discs. We find that two-fluid effects are most important for grains larger than 1~$\mu$m, and that the peak dust temperature within an accretion shock provides a signature of the dust-to-gas ratio of the infalling material.
\end{abstract}
\begin{keywords}
dust -- shock waves -- protoplanetary disc -- accretion
\end{keywords}
\section{INTRODUCTION }
\label{sec:intro}

Dust is found to play a key role in numerous astrophysical evironments. In the interstellar medium (ISM), dust is heavily involved in controlling the thermodynamics by being a major coolant, collisional partner and source of opacity. It allows us to probe the magnetic field by measuring the polarization in thermal dust emission \citep{planck_collaboration_planck_2016}, and is crucial to the formation of H$_2$ in molecular clouds by providing a catalytic surface and by attenuating the dissociating ultraviolet radiation field \citep{glover_is_2012}. Thermal dust emission is observed with telescopes such as Spitzer \citep[e.g.][]{stephens_spitzer_2014}, Herschel \citep[e.g.][]{launhardt_earliest_2013}, and ALMA \citep[e.g.][]{alma_partnership_2014_2015}, and these observations are used to obtain properties of the gas. Thus it is crucially important not only to understand the properties of dust grains, but also their coupled evolution with the gas phase in order to rigorously relate dust emission to properties of the ISM.

The importance of coupled gas-dust modelling has motivated the development of numerical codes designed to simulate gas and dust in various astrophysical systems. For example, radial gradients of gas pressure in protoplanetary discs induce dust clumping leading to planetesimal formation \citep{bai_effect_2010}, large dust-to-gas variations occur in turbulent molecular clouds and dust filaments do not necessarily correlate with gas filaments \citep{hopkins_fundamentally_2016}, and dust-gaps are cleared more easily than gas-gaps in protoplanetary discs \citep{paardekooper_dust_2006}. Both grid-based \citep{bai_particle-gas_2010} and particle-based smoothed particle hydrodynamics codes \citep{laibe_dusty_2012} have been used to study dusty gas flows in the ISM. However, \cite{laibe_dustybox_2011} highlight a lack of simple analytic solutions to benchmark dusty gas codes in astrophysical conditions. The main goal of this work is to provide such a simple solution by computing the structure of steady-state, planar two-fluid dusty gas shock waves. Unlike the standard shock-tube tests, steady shocks comprise only one hydrodynamic component with a structure that can be computed by simply integrating the governing ordinary differential equations, as we do in Section~\ref{dustyshocks-benchmark}. The numerical simulation is also simple: drive a piston represented by reflective boundary conditions into a uniform medium. This simple test can be used to benchmark how numerical codes behave with different dust-to-gas ratios, or e.g. linear, quadratic, or Epstein forms of the drag coefficients.

Two-fluid dusty shocks are not just ideal benchmarks for numerical codes. Supersonic flows occur ubiquitously in astrophysical systems. For example, in the inside-out collapse model of protoplanetary cores, material becomes thermally unsupported and free-falls onto the protoplanetary disc at a few km/s. The sound speed in the gas is only $\sim$0.2 km/s, and so a shock wave forms as the material decelerates to settle onto the disc. In Section~\ref{sec:dustyshocks-accretionshock} we provide an application of our two-fluid shock solutions to study this type of accretion shock.

\begin{figure*}
  \centering
    \includegraphics[width=0.7\textwidth]{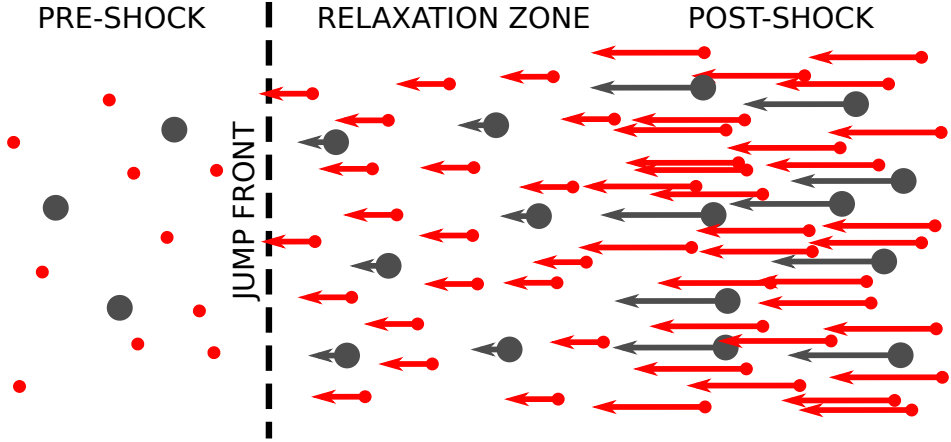}
    \caption{Sketch of a J-type dust-gas two-fluid shock in the frame of reference where the preshock fluid is stationary. The red, smaller circles refer to gas particles whereas the larger gray circles refer to dust particles. The black dashed line marks the jump transition in the gas fluid that takes place on the order of a few mean free paths.}
  \label{fig:cartoon}
\end{figure*}

\section{THEORY}\label{sec:dustyshocks-theory}

In this section we outline the set of equations that describe the two-fluid dust-gas system. We use these equations to derive the dispersion relation for linear waves in the combined fluid, and discuss how this affects the possible dust-gas shock structures.

\subsection{Fluid Equations} 
For a fluid with gas density $\rho$, velocity $\mathbf{v}$ and pressure $P$, and dust density $\rho_d$ and velocity $\mathbf{v}_d$, the equations of continuity and conservation of momentum for the gas can be written
\begin{align}
\label{eq:cont_gas} & \frac{\partial \rho}{\partial t} + \nabla \cdot \left( \rho  \mathbf{v} \right) = 0,\\
\label{eq:mom_gas} & \rho\frac{\partial \mathbf{v}}{\partial t} + \rho \left( \mathbf{v} \cdot \nabla \right) \mathbf{v}= - \nabla P - \mathbf{F}_\mathrm{drag},
\end{align}
where $\mathbf{F}_\mathrm{drag}$ is the rate at which momentum is added to the gas via drag from the dust fluid. The analogous equations for the dust fluid are
\begin{align}
\label{eq:cont_dust} & \frac{\partial \rho_d}{\partial t} + \nabla \cdot \left( \rho_d  \mathbf{v}_d \right) = 0, \\
\label{eq:mom_dust}  & \rho_d\frac{\partial \mathbf{v}_d}{\partial t} + \rho_d \left( \mathbf{v}_d \cdot \nabla \right) \mathbf{v}_d = \mathbf{F}_\mathrm{drag}, 
\end{align}
where we have assumed the dust to be pressureless. We close the fluid equations with a polytropic equation of state
\begin{align*}
P = c_s^2 \rho ^\gamma
\end{align*}
where $c_s$ is the sound speed.

The drag term has been thoroughly discussed by \cite{laibe_dusty_2012-1} for various forms of astrophysical interest. It is generally proportional to a power law of the velocity drift between the two fluids. If we consider the linear drag regime where the drag term on the total fluid can be written as
\begin{align*}
\mathbf{F}_\mathrm{drag} = K \left(\mathbf{v} - \mathbf{v}_d \right)
\end{align*}
for drag coefficient $K$, then linear waves obey the dispersion relation given by
\begin{align}
\omega\left(\omega^2 -  k^2 c_s^2\right) + iK\left(\rho_0^{-1} + \rho_{d0}^{-1}\right)\left( \omega^2 - k^2 \tilde{c}_s^2\right) = 0 \label{eq:dispersion}
\end{align}
where $\rho_0$ and  $\rho_{d0}$ are the unperturbed gas and dust density, respectively, and
\begin{align*}
\tilde{c}_s = c_s \left(1 + D\right)^{-1/2}
\end{align*}
with the dust to gas ratio
\begin{align*}
D = \frac{\rho_{d0}}{\rho_0}.
\end{align*}
In the limit of weak coupling between the dust and gas ($K \to 0$) we recover the dispersion relation for ordinary sound waves in a gas with phase velocity $\omega / k = c_s$. In the strong coupling limit ($K \to \infty$) the complex term of equation~\eqref{eq:dispersion} dominates, and so waves travel at the combined sound speed $\tilde{c}_s$.

The two signal speeds in the system, $c_s$ and $\tilde{c}_s$, determine the possible structures of dust-gas shocks. The dust-gas mixture will behave as a single fluid far ahead and behind a shock and so the combined sound speed $\tilde{c}_s$ is the relevant signal speed that, in the frame of reference comoving with the shock, the fluid velocity must transition across. As $\tilde{c}_s$ is necessarily less than $c_s$, we will see that two distinct classes of shocks arise depending on whether the shock speed is greater or less than the gas signal speed $c_s$.

For a supersonic shock (shock velocity $v_s > c_s$ ) the preshock fluid is overrun by high density gas in a thin shock front a few mean free paths wide that resembles an ordinary gas dynamic shock. The dust particles cannot respond quickly, and so there is a relaxation zone wherein the dust particles are accelerated until the two fluids flow at the same velocity. This structure is qualitatively sketched in Figure~\ref{fig:cartoon}.

When the shock speed is between the two signal speeds, sound waves in the gas fluid can travel ahead of the shock front and compress the gas and dust in such a way that all the fluid variables remain continuous through the shock. We will call these two classes J-type and C-type shocks in analogy to the kinds of magnetised two-fluid shocks outlined by \cite{draine_multicomponent_1986}, where ion-magnetosonic waves can travel ahead of a jump front to form a ``magnetic precursor". Section~\ref{sec:dustyshocks-statpoints} characterises these classes in further detail.

\section{STEADY PLANAR SHOCKS}\label{dustyshocks-benchmark}

In this section we describe the first order differential equation that we solve to investigate the structure of two-fluid dust-gas shocks. We characterise the initial stationary states to outline the criteria for J- and C-type shocks to occur and discuss their structure. These shock solutions are ideal tests for benchmarking numerical codes that wish to simulate dusty gas. Our \textsc{python} code that returns the shock solutions described in this section is publicly available on the Python Package Index\footnote{https://pypi.python.org/pypi/DustyShock} and BitBucket\footnote{https://bitbucket.org/AndrewLehmann/dustyshock}.

In the standard shock-tube problem \citep{sod_survey_1978} the simple setup breaks up into a shock wave, rarefaction wave and a contact discontinuity. In the two-fluid dust-gas version of this problem there is no known analytic solution, but \cite{saito_numerical_2003} find that a steady-state shock solution fits one of the components well. Unlike the Sod shock-tube, the steady shocks we compute here are very simple, comprising just one hydrodynamic structure. A numerical code can then be tested against the steady solution by setting up a reflective boundary representing a piston as the driver of the shock, as in \cite{toth_numerical_1994}.

\subsection{Numerical Integration} \label{sec:dustyshocks-shockeqs}
Assuming a steady state, one-dimensional structure varying in the $z$-direction and a power law drag term, the gas fluid equations reduce to
\begin{align}
\label{eq:cont_g} & \frac{d}{d z} \left( \rho  v \right) = 0\\
\label{eq:momz_g} &\frac{d}{d z} \left( \rho  v^2 + c_s^2 \rho \right)= K \left| v_d - v \right|^n
\end{align}
and the dust fluid equations reduce to
\begin{align}
& \frac{d}{d z} \left(\rho_d v_d \right) = 0 \label{eq:cont_d} \\
& \frac{d}{d z} \left( \rho_d  v_d^2 \right) = -K \left| v_d - v \right|^n. \label{eq:momz_d} 
\end{align}
We get the dimensionless derivative of the dust velocity from combining equation~\eqref{eq:cont_d} and \eqref{eq:momz_d}:
\begin{align}
\frac{d w_d}{d\xi} = -\left|w_d - w\right|^n \label{eq:ode}
\end{align}
for the normalised velocities and position defined by
\begin{align}
w_d &= \frac{v_d}{v_s}\nonumber\\
w &= \frac{v}{v_s}\nonumber\\
\xi & = \frac{K}{\rho_{d0} v_s^{2-n}} z \label{eq:scaling}
\end{align}
where $v_s$ is the shock velocity and $\rho_{d0}$ is the preshock dust mass density. The normalised gas velocity $w$ can be expressed in terms of $w_d$ as a solution of the quadratic equation
\begin{align}\label{eq:root}
w^2 + \left[ D\left( w_d - 1\right) -1 - \mathcal{M}^{-2}\right] w  + \mathcal{M}^{-2} = 0
\end{align}
where we have used the sonic Mach number $\mathcal{M} \equiv v_s/c_s$, assuming that initially the gas and dust flow together at the shock velocity $v_s$. The two roots of the quadratic represent supersonic and subsonic (with regards to $c_s$) gas velocities.

The fluid variables defining the shock structure are obtained by integrating the first order ordinary differential equation (ODE) defined by equation~\eqref{eq:ode}. In this paper we use the open source \textsc{python} module \emph{scikits.odes}\footnote{https://github.com/bmcage/odes}. This module solves initial value problems for ODEs using variable-order, variable-step, multistep methods.

\subsection{Stationary Points}\label{sec:dustyshocks-statpoints}

The pre-shock state, defined by the gas density $\rho_0$, dust density $\rho_{d0}$, shock velocity $v_s$ and initial temperature $T_0$ is a stationary point. In order to classify this state, we perturb the initial state and let the perturbation grow (or decay) exponentially as follows:
\begin{align*}
& v = v_s + \delta v e^{\lambda z} \\
& v_d = v_s + \delta v_d e^{\lambda z} \\
& \rho = \rho_0 + \delta \rho e^{\lambda z} \\
& \rho_d = \rho_{d0} + \delta \rho_d e^{\lambda z}\\
&P = \rho_{g0} c_s^2 + c_s^2\delta \rho e^{\lambda z}
\end{align*}
where we have assumed isothermality. Substituting these perturbed variables back into equations~\eqref{eq:cont_g}-\eqref{eq:momz_d} gives the eigenvalue
\begin{align}
\lambda = - \frac{K}{\rho_{d0} v_s} \frac{1 + D}{D } \frac{v_s^2 - \tilde{c}_s^2}{v_s^2 - c_s^2} \label{eq:eigenvalue}
\end{align}
This expression implies that for supersonic shocks ($v_s > c_s > \tilde{c}_s$), the eigenvalue $\lambda < 0$ and hence the perturbation decays. That is, the initial state is a stable stationary point. Hence it requires an initial discontinuity (jump) to get across the sound speed. In this kind of shock, the gas fluid is highly compressed over a few mean free paths. The dust cannot respond quickly, and so this jump is determined by the hydrodynamic jump conditions. For the isothermal shocks computed here, the density jumps as
\begin{align*}
\frac{\rho_2}{\rho_1} = \mathcal{M}^2
\end{align*}
while the velocity switches from the supersonic to the subsonic root using equation~\eqref{eq:root}. 

By replacing $v_s$ with the post-shock solution $v_\mathrm{post}$ we can explore the final state. A shock in the total gas-dust fluid is a transition across the total speed $\tilde{c}_s$, and so $v_\mathrm{post} < \tilde{c}_s$. Thus the eigenvalue near the post-shock state $\lambda <0$, which defines a stable point, so jumping near this state will settle onto it. 
Two-fluid dust-gas J-type shocks were first discussed by \cite{carrier_shock_1958}, and have been thoroughly studied for various non-astrophysical applications \citep[see review by][and references therein]{igra_dusty_1988}.

\subsubsection*{J-type shock solutions}

\begin{figure}
  \centering
    \includegraphics[width=\columnwidth]{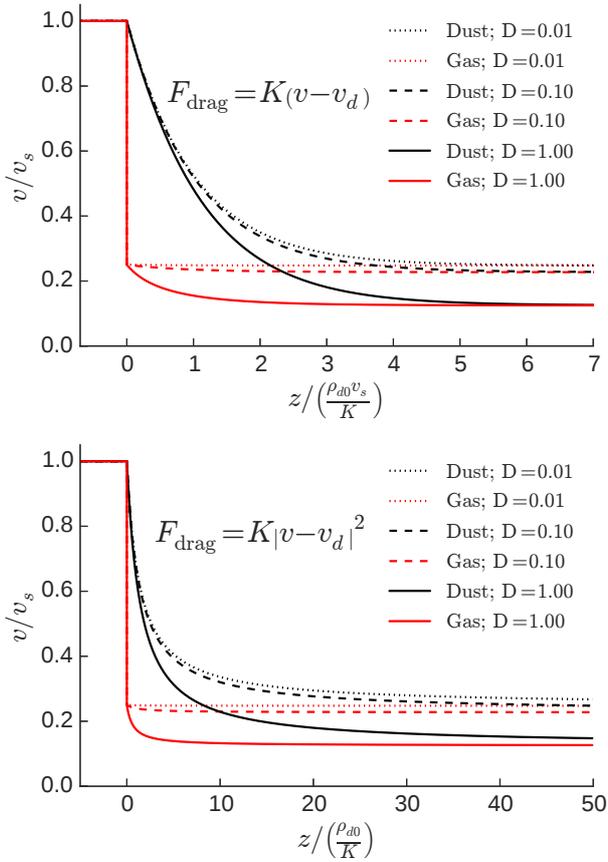}
    \caption{J-type gas-dust shocks with Mach number $\mathcal{M}=2$ and initial dust-to-gas ratios $D=1$ (solid lines), $D=0.1$ (dashed lines) and $D=0.01$ (dotted lines). The red lines give the gas velocity and the black lines gives the dust velocity, both normalised to the shock velocity. The upper panel are solutions computed with a linear drag term while the lower panel uses a quadratic drag term. Note that the $z$-scale normalisation differs according to eq.~\eqref{eq:scaling}.}
  \label{fig:jshock}
\end{figure}

An example of the velocity structure of an isothermal J-type shock in the frame of reference comoving with the shock is shown in fig.~\ref{fig:jshock}. The upper panel shows shock structures computed with Mach number $\mathcal{M}=2.0$ using a linear drag term and initial dust-to-gas ratios varying from 0.01--1. After the initial hydrodynamic jump in the gas particles, the dust lags behind the gas, but both eventually settle to the same velocity (below $\tilde{c}_s$). The combined fluid velocities far ahead ($v_s$) and behind ($v_2$) the shock front are related by
\begin{align}\label{eq:vjump}
\frac{v_2}{v_s} = \left(\frac{\tilde{c_s}}{v_s}\right)^2 = \left(1+D\right)^{-1}\mathcal{M}^{-2}.
\end{align}
Hence reducing $D$ changes the shock structure by increasing the post-shock velocity that the solution settles to. This effect allows this steady state solution to test how a numerical code behaves with different dust-to-gas ratios.

The lower panel of fig.~\ref{fig:jshock} shows isothermal shock solutions with the same conditions as the upper panel except that they are computed using a quadratic drag term
\begin{align*}
F_{\mathrm{drag}} = K \left| v - v_d \right|^2.
\end{align*}
The structure is qualitatively similar to the linear drag case, however the shock thickness is an order of magntitude larger (in the dimensionless position variable $\xi$). From eq.~\eqref{eq:ode}, the shock thickness $\Delta \xi \sim \left(\Delta w\right)^{1-n}$, and $\Delta w$ is necessarily between 0 and 1. Hence the the shock thickness increases with the index of the power-law drag. For this reason, this solution also has an extended tail of very small but finite $\Delta w$. These stark differences allow a numerical code's implementation of different drag coefficients to be tested by the shock problem. Note that regardless of the form of the drag term, the shock solution settles onto the same post-shock velocity. This is because the jump conditions relate the velocities of the \textit{combined} fluid on either side of the shock.

The different shock structures resulting from different drag terms could be used to test a numerical code's implementation of the drag. However, in many codes artificial viscosity is used to smear a discontinuity over a finite distance. The scaling factor
\begin{align*}
\frac{K}{\rho_{d0}v_s^{2-n}}
\end{align*} 
can be adjusted until the relaxation tail of the J-type analytic shock solution spreads over many computational cells, so that the behaviour of a numerical code in certain regimes of the drag coefficient (small $K$) and/or dust density (large $\rho_{d0}$) can be tested. The C-type shock solution in the following section avoids this limitation because it lacks any discontinuity. The structure also depends on the dust-gas interaction more crucially than in J-type shocks, and may therefore be a more appropriate test problem for dusty gas numerical codes.

\subsubsection*{C-type shock solutions}

\begin{figure}
  \centering
    \includegraphics[width=\columnwidth]{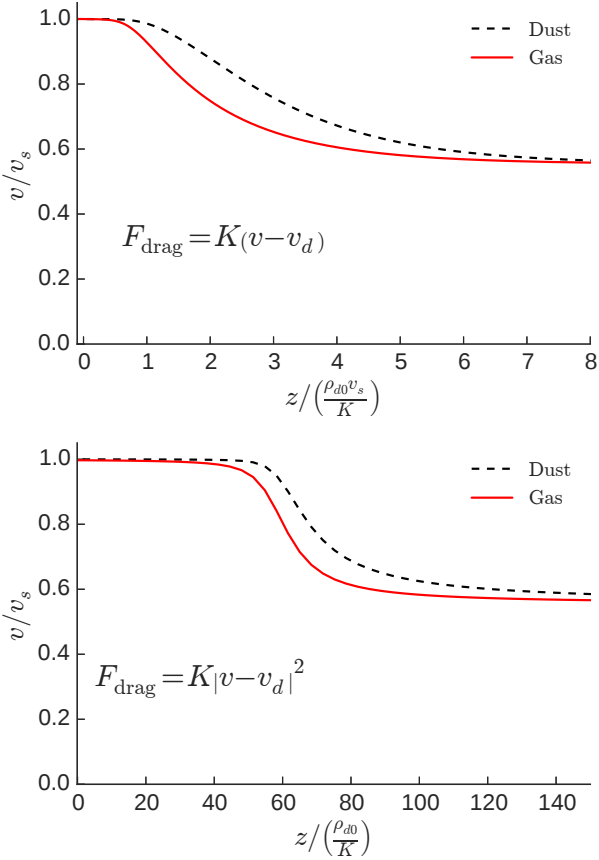}
    \caption{Same as Fig.~\ref{fig:jshock} but for a C-type shock with Mach number $\mathcal{M}=0.95$ and $D=1$.}
  \label{fig:cshock}
\end{figure}

For C-type shock solutions to exist we require a positive eigenvalue in equation~\eqref{eq:eigenvalue}, so that perturbations smoothly grow away from the preshock state. The shock speed must then be in the range
\begin{align*}
\tilde{c}_s < v_s < c_s
\end{align*}
and hence the Mach number is necessarily below unity for this type of shock. Without cooling, this shock will smoothly settle onto the post-shock state. However, with cooling there is the potential for the gas sound speed to drop (as the gas compresses in the shock) faster than the velocity. If $c_s = v$ in the shock somewhere, a jump will be required. Otherwise, the fluid variables will remain continuous throughout the shock. This kind of shock was first investigated by \cite{kriebel_analysis_1964} and further developed by \cite{miura_weak_1972}.

An example of the velocity structure of isothermal C-type shocks is shown in fig.~\ref{fig:cshock}. The pre-shock variables are the same as for the J-type shock discussed above, except that the Mach number $\mathcal{M}=0.95$. The structure is a factor of $\sim$2 wider than the example J-type shock, and as the two fluids smoothly settle onto the post-shock state the drift velocity remains small. 

When computed with a quadratic drag term (lower panel), the solution initially changes slowly when the drift velocity is small and the drag is smaller than the linear regime. When the drift is large the velocity changes rapidly until the drift is small again. Thus the structure is much thinner than the linear drag solution in the centre of the shock when the drift is large (note the $z$-scale differs by a factor of $10^3$ in the lower panel), but takes a long time for the two fluids to come to rest with respect to each other.

Recall that the shock velocity $v_s$ for C-type dusty gas shock is restricted to the range
\begin{align*}
\tilde{c}_s < v_s < c_s,
\end{align*}
and that the combined fluid speed is related to the pure gas sound speed by the factor
\begin{align*}
\left(1+D\right)^{-1/2}.
\end{align*}
This means that when the initial dust-to-gas ratio becomes small---such as the typical interstellar value of 0.01---$v_s$ cannot be much larger than $\tilde{c}_s$, resulting in a very weak shock. For this reason, C-type dusty shocks may not relevant to the general ISM. However, strong variations of $D$ have been found in simulations of turbulent molecular clouds when dust-gas decoupling has been modeled \citep{hopkins_fundamentally_2016}, and values as high as unity have been used in protostellar discs \citep[e.g.][]{dipierro_planet_2015} to account for dust migration to the inner parts of the disc.

We have presented two potential benchmarking problems for numerical codes seeking to simulate dusty gas systems. These problems allow a code to test the implementation of different forms of drag, the strength of the drag (K) and the dust to gas mass density ratio (D). In the following section we provide an astrophysical application of two-fluid dust-gas shocks.

\section{PROTOPLANETARY DISC ACCRETION SHOCK}\label{sec:dustyshocks-accretionshock}
Here we present an astrophysical application of two-fluid dust-gas shocks. At very early stages of star formation, the system is characterised by an embedded protostellar disc surrounded by an infalling envelope of dust and gas. The material falls, due to gravity, through an accretion shock and then eventually settles onto the disc. We study this accretion shock by modeling it as a two-fluid dust-gas shock.

\subsection{Shock Parameters}\label{sec:dustyshocks-params}

The accretion rate onto a protoplanetary disc in the inside-out collapse model of a singular isothermal sphere can be approximated \citep{larson_physics_2003} as
\begin{align*}
\dot{M} \sim \frac{c_s^3}{G} = 2\times 10^{-6} \, \mathrm{M_\odot}\,\mathrm{yr}^{-1}
\end{align*}
where $G$ is the gravitational constant, and we have assumed the typical interstellar medium sound speed $c_s=0.2$~km~s$^{-1}$. This mass is spread over the area of the disc, so that the mass flux of the accretion shock is
\begin{align}
\rho_0 v_s &= \frac{\dot{M}}{\pi R_d^2} \sim \frac{c_s^3}{G\pi R_d^2}\\ 
&\sim 1.7 \times 10^{-11} \left( \frac{R_d}{100 \, \rm{au}}\right)^{-2} \, \rm{g/s/cm}^2\label{eq:mass_flux}
\end{align}
where $R_d$ is the disc radius. The material accretes in free-fall and thus reaches the protoplanetary disc at the escape speed. The gas meets a disc in Keplerian motion, and so the shock velocity
\begin{align}\label{eq:shockvel}
v_s \sim \sqrt{\frac{G M_*}{r}} 
\end{align}
where $r$ is the distance from the central protostar with mass $M_*$. Using this velocity and eq.~\eqref{eq:mass_flux} we get a preshock density of
\begin{align*}
\rho_0 \sim 4 \times 10^{-17} \left( \frac{R_d}{100 \, \mathrm{au}} \right)^{-2}\left(\frac{r}{50\,\mathrm{au}}\right)^{1/2} \, \rm{g/cm}^3.
\end{align*}
This corresponds to a total hydrogen number density, $n_\mathrm{H} \approx n(\mathrm{H}) + 2n(\mathrm{H}_2)$, of
\begin{align*}
n_0 = \frac{\rho_0}{1.4 m_\mathrm{H}} \sim 2\times 10^7 \left( \frac{R_d}{100 \, \mathrm{au}} \right)^{-2} \left(\frac{r}{50\,\mathrm{au}}\right)^{1/2} \, \mathrm{cm}^{-3}.
\end{align*}

\begin{figure}
  \centering
    \includegraphics[width=\columnwidth]{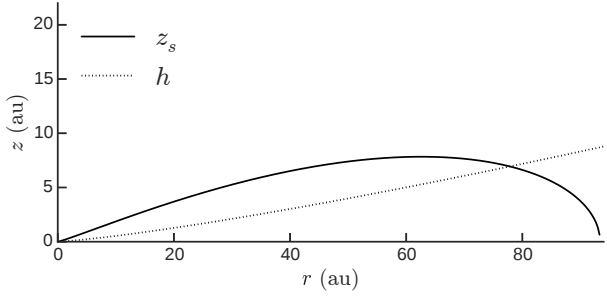}
    \caption{Location of shock (solid line), where the ram pressure of free-falling material balances the thermal pressure of the protoplanetary disc. The dotted line marks the scale height for comparison.}
  \label{fig:shock-height}
\end{figure}

\begin{figure}
  \centering
    \includegraphics[width=\columnwidth]{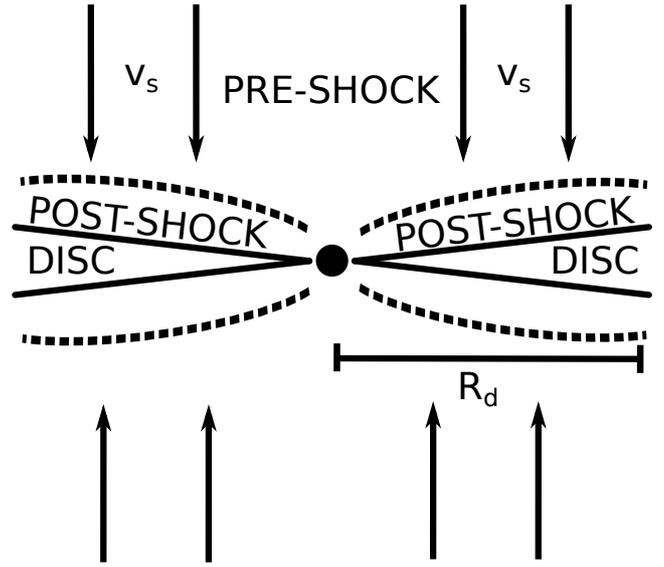}
    \caption{Sketch of model for accretion shock (dashed) onto protoplanetary disc.}
  \label{fig:accretion_sketch}
\end{figure}

The shock will be located where its ram pressure is balanced by the thermal pressure of the disc. That is, where
\begin{align}
n_0 v_s^2 = n_d(r,z) c_s(r)^2 \label{eq:rambalance}
\end{align}
where the disc density and sound speed are functions of radial distance from the star, $r$, and vertical distance from the disc, $z$. These functions can be approximated in the minimum mass solar nebula model \citep{wardle_magnetic_2007} as
\begin{align*}
n_d(r,z) &\sim 5.8\times 10^{14} \, \mathrm{cm}^{-3} \left( \frac{r}{\mathrm{au}} \right)^{-11/4} \exp \left( -\frac{z^2}{2h^2} \right)\\
c_s(r) &\sim 0.99 \, \mathrm{km \, s}^{-1} \left( \frac{r}{\mathrm{au}} \right)^{-1/4}
\end{align*}
with the scale height, $h$, is given by
\begin{align*}
\frac{h}{r} \sim 0.03 \left( \frac{r}{\mathrm{au}} \right)^{1/4}.
\end{align*}
Subsituting these approximations into equation~\eqref{eq:rambalance} and rearranging for the vertical height gives
\begin{align*}
z_s^2 = 2 h^2 \ln \left( 2.62\times 10^5  \left( \frac{R_d}{100 \, \mathrm{au}} \right)^2 \left( \frac{r}{\mathrm{au}} \right)^{-11/4}\right) .
\end{align*}
The location where the shock ram pressure balances the disc thermal pressure ($z_s$) is shown in fig.~\ref{fig:shock-height} for a disc radius $R_d=100$~au. Beyond $r\sim 95$~au, the disc density and sound speed has dropped so low that its thermal pressure never balances the shock ram pressure. In this region the shock will run up against the material freely falling from the other side of the disc. A sketch of the system is shown in fig.~\ref{fig:accretion_sketch}. In the next section we model this accretion shock using representative values of the shock velocity and preshock density between 30--80~au, i.e. $\sim$4.0~km~s$^{-1}$ and $\sim$4$\times 10^{-17}$~g~cm$^{-3}$, respectively.

\subsection{Physical Processes}

Following \cite{draine_multicomponent_1986} the drag term, or momentum rate of change change per volume due to elastic scattering, can be expressed as
\begin{align*}
F = \frac{\sigma \rho \rho_d}{m_d} \sqrt\frac{2kT}{\pi \mu} \left( v_d - v \right) I\left(v, v_d, T \right)
\end{align*}
where $\sigma$ is the dust cross-section, $m_d$ is the mass of a dust particle, $\mu$ is the mean mass per gas particle ($(7/3)m_\mathrm{H}$ in molecular gas), $T$ is the gas temperature and $k$ is the Boltzmann constant. The function $I(v, v_d, T)$ is well approximated by
\begin{align*}
I(v, v_d, T) \approx \frac{8}{3}\left( 1 + \frac{9 \pi}{64} \frac{\frac{1}{2}\mu \left| v- v_d \right|^2}{kT}\right)^{1/2}.
\end{align*}
The derivative of the dust velocity through the shock then becomes
\begin{align}\label{eq:ode2}
\frac{d v_d}{dz} = \alpha v_s \sqrt{\frac{kT}{\mu}} I(v, v_d, T) \frac{v - v_d}{v_d v} 
\end{align}
where
\begin{align*}
\alpha =  \frac{\rho_0 r_d^2 \sqrt{2\pi}}{m_d}
\end{align*}
for preshock gas density $\rho_0$ and dust radius $r_d$. We have assumed that the dust cross-section is its circular area.

For the $v_s \sim 4$~km~s$^{-1}$ shock we are considering, the gas temperature immediately behind the shock will jump from 10 to $\sim$10$^3$~K. If we estimate the drift velocity $\Delta v = v_d- v \sim v_s/2$, the post-jump gas velocity $v\sim v_s/5$, assume spherical carbon dust particles with homogeneous density $\sim$2.2~g~cm$^{-3}$, then we can estimate the shock thickness to be
\begin{align*}
\Delta z \sim \frac{v_d v}{\alpha v_s I} \left(\frac{kT}{\mu}\right)^{-1/2} \sim 0.06 \, \mathrm{au} \left( \frac{r_d}{\mu\mathrm{m}}\right).
\end{align*}
Hence the shock remains thin compared to the size of the system shown in fig.~\ref{fig:shock-height}.

In order to include heating and cooling we use the energy equation for the gas fluid
\begin{align*}
v\frac{dP}{dz} + \gamma P \frac{dv}{dz} = \left(\gamma-1\right) \left(\Gamma - \Lambda\right)
\end{align*}
where $\Gamma$ is the heating rate per unit volume and $\Lambda$ is the cooling rate per unit volume. Combined with an ideal equation of state $P=nkT$, we derive the gas temperature derivative
\begin{align*}
\frac{1}{T}\frac{dT}{dz} = \frac{1}{\rho_0 v_s^3} \frac{\gamma-1}{w^2 -\gamma\tau} \left(\left(\frac{w^2}{\tau} - 1 \right) \left(\Gamma - \Lambda\right) + v F \right)
\end{align*}
where $\tau=kT/\mu v_s^2$. We also modify equation~\eqref{eq:root} to account for the variable temperature
\begin{align*}
w^2 + \left[ D\left( w_d - 1\right) -1 - \mathcal{M}^{-2}\right] w  + \frac{kT}{\mu v_s^2} = 0.
\end{align*}
Finally, the temperature jumps from $T_1$ to $T_2$ across the initial discontinuity following
\begin{align*}
\frac{T_2}{T_1} &= \left( 1 + \frac{2 \gamma}{\gamma + 1}\left( \mathcal{M}^2 -1 \right) \right) \frac{\mathcal{M}^2 \left( \gamma -1 \right) +2}{\mathcal{M}^2\left( \gamma +1 \right)}.
\end{align*}

From \cite{draine_multicomponent_1986}, the rate of change per volume of the thermal energy content of gas due to elastic scattering by dust with a velocity-independant cross section, $\sigma$, is
\begin{align*}
\Gamma_\mathrm{drag} = \frac{\sigma \rho \rho_d}{m_d^2} \sqrt{\frac{8 k T}{\pi \mu}}  \left[ k\left( T - T_d \right)I_2 + kT_d I_3\right]
\end{align*}
where
\begin{align*}
I_2 &\approx \left( 1 + \frac{9 \pi}{64}\frac{\frac{1}{2}\mu \left| v- v_d \right|^2}{kT} \right)^{1/2} \left( 4 + \frac{8}{3}\frac{\frac{1}{2}\mu \left| v- v_d \right|^2}{kT} \right)
\end{align*}
and
\begin{align*}
I_3 &\approx \left( 1 + \frac{9 \pi}{64}\frac{\frac{1}{2}\mu \left| v- v_d \right|^2}{kT} \right)^{1/2} \frac{8}{3}\frac{\frac{1}{2}\mu \left| v- v_d \right|^2}{kT}.
\end{align*}

To calculate the dust temperature we assume that the frictional heating per grain, $\Gamma_\mathrm{drag}/n_d$, is always balanced by the power radiated by a dust grain. Following \cite{draine_physics_2011}, grains lose energy by infrared emission at a rate, per grain,
\begin{align}\label{eq:L_grain}
\Lambda _d = 4 \pi r_d^2 \left\langle Q_\mathrm{abs} \right\rangle \sigma T_d^4
\end{align}
where $\sigma$ is the Stefan-Boltzmann constant and $\left\langle Q_\mathrm{abs} \right\rangle$ is the Planck-averaged emission efficiency, which for carbon grains is
\begin{align*}
\left\langle Q_\mathrm{abs} \right\rangle_\mathrm{C} \sim 8 \times 10^{-7} \left( \frac{r_d}{0.1 \mu\mathrm{m}} \right) \left( \frac{T_d}{\mathrm{K}} \right)^2.
\end{align*}

We include rotational line cooling from CO and H$_2$ using the cooling functions of \cite{neufeld_radiative_1993} and \cite{neufeld_thermal_1995}. They give the cooling rate per volume $\Lambda (M) = n(M)n(\mathrm{H}_2)L$ for a molecule $M$ using a cooling rate coefficient $L$ obtained by fitting to four parameters of the form
\begin{align*}
\frac{1}{L_\text{M}} = \frac{1}{L_0} + \frac{ n(\text{H}_2) }{ L_\mathrm{LTE} } + \frac{1}{L_0}\left( \frac{ n(\text{H}_2) }{ n_{1/2} }  \right) ^\alpha \left( 1 -  \frac{ n_{1/2} L_0 }{ L_\mathrm{LTE}} \right).
\end{align*}
The parameters $L_0$, $L_\mathrm{LTE}$, $n_{1/2}$ and $\alpha$ are tabulated for temperatures up to a few thousand K, and depend on an optical depth parameter $\tilde{N}$. Modeling the shock as a plane-parallel slab of thickness $d$, this parameter is given as
\begin{align*}
\tilde{N}(\mathrm{CO}) = \frac{n(\mathrm{CO})d}{9\Delta v}
\end{align*}
We have chosen a CO abundance $x(\mathrm{CO})=1.24\times 10^{-4}$ with respect to the total hydrogen density and molecular hydrogen abundance $x(\mathrm{H}_2)=0.5$, both constant throughout the shock. We thus use an optical depth parameter $\tilde{N}(\mathrm{CO})\sim 10^{15}$~cm$^{-2}$/(km/s), appropriate for a shock thickness of 1~au and $\Delta v$=4~km~s$^{-1}$.

Finally, we choose a preshock gas and dust temperature $T_{g0}=T_{d0}=10$~K, corresponding to an isothermal sound speed $c_s=0.188$~km~s$^{-1}$, and we do not allow either temperature to fall below their initial value.

\subsection{Results and Discussion}

\begin{figure}
  \centering
    \includegraphics[width=\columnwidth]{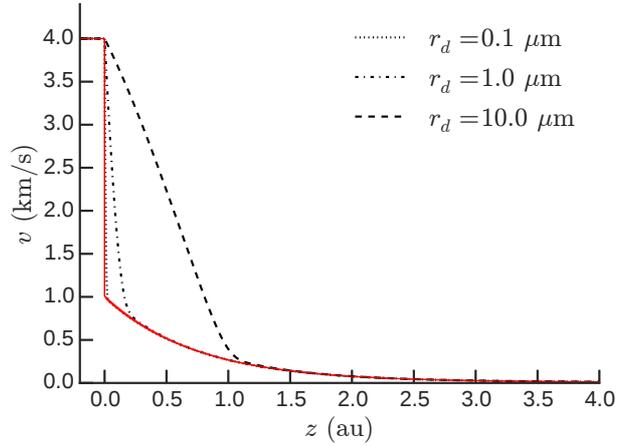}
    \caption{Accretion shock velocity profile for $v_s$=4~km/s, $\rho_0=4\times 10^{-17}$~g/cm$^3$, and preshock dust-to-gas ratio $D=0.01$. Dust velocities for three different dust sizes are shown, $r_d$=0.1 (dotted), 1 (dash-dotted) and 10 (dashed)~$\mu$m. The gas velocity profiles (solid red line) cannot be distinguished in the three cases.}
  \label{fig:accretion_profile_size}
\end{figure}

With the heating and cooling processes in place, we numerically integrate the coupled ODEs as discussed in \ref{sec:dustyshocks-statpoints} for J-type shocks. We first investigate the effect of different sizes of dust grains by considering a constant initial dust-to-gas ratio $D=0.01$. The resulting velocity profiles computed for dust sizes $r_d=0.1$, 1 and 10~$\mu$m are shown in fig.~\ref{fig:accretion_profile_size}. 

When $r_d=0.1$~$\mu$m, the region of the shock with any drift between the dust and gas velocities is negligible compared to the size of the shock, and so closely resembles a one-fluid shock. However, when $r_d=10$~$\mu$m this region is about half the size of the shock. Hence we expect two-fluid effects to be more prominent in accretion shocks where the dust has coagulated into large grains ($r_d > 1$~$\mu$m). Note that the shock thickness $\sim$4~au, which is approaching the limits of validity for this model.

\begin{figure}
  \centering
    \includegraphics[width=\columnwidth]{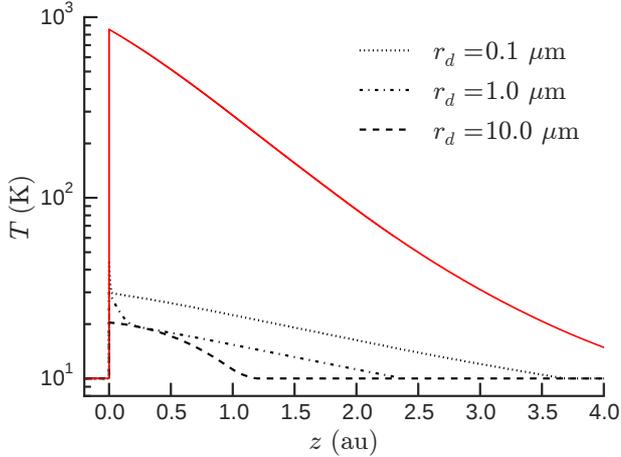}
    \caption{Accretion shock temperature profile for $v_s$=4~km/s, $\rho_0=4\times 10^{-17}$~g/cm$^3$, and preshock dust-to-gas ratio $D=0.01$. Dust temperatures for three different dust sizes are shown, $r_d$=0.1 (dotted), 1 (dash-dotted) and 10 (dashed)~$\mu$m. The gas temperature profiles (solid red line) cannot be distinguished in the three cases.}
  \label{fig:accretion_profile_temps}
\end{figure}

The heating and cooling of the dust depends on grain radius such that larger grains will reach lower temperatures. The temperature profiles of the gas and dust for the same shocks shown in fig.~\ref{fig:accretion_profile_size} are shown in fig.~\ref{fig:accretion_profile_temps}. The peak dust temperature decreases from 40 to 20~K with increasing grain radius. In addition, the smaller grains take longer to cool, retaining their increased temperature for a larger fraction of the shock.

\begin{figure}
  \centering
    \includegraphics[width=\columnwidth]{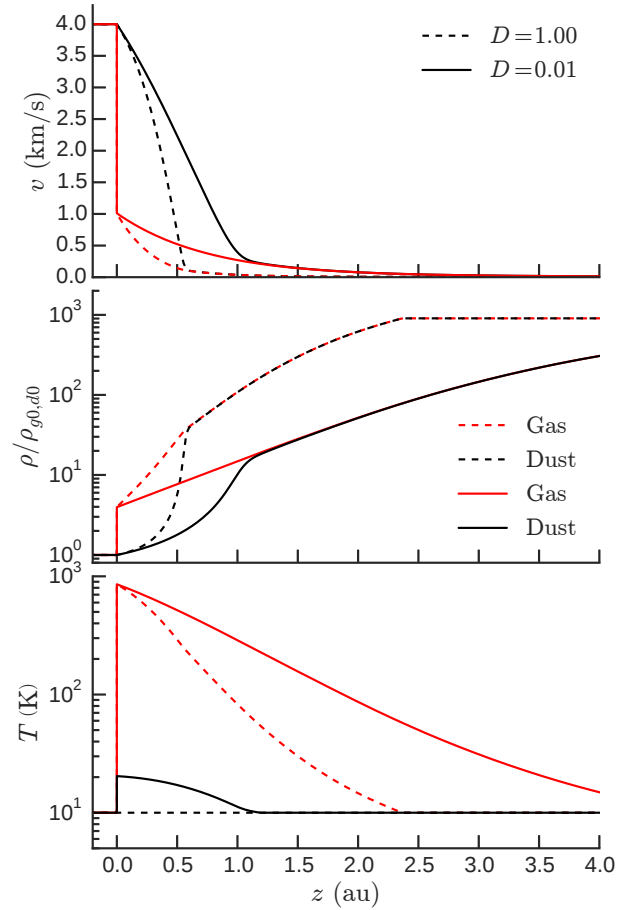}
    \caption{Accretion shock profile for $v_s$=4~km/s, $\rho_0=4\times 10^{-17}$~g/cm$^3$, and dust size $r_d=10$~$\mu$m. Upper panel shows the velocity profiles, middle panel shows the density profiles, and the lower panel shows the temperature profiles. Solid lines are shocks with initial dust-to-gas ratio $D=0.01$ while dashed lines have $D=1$. Black lines refer to dust variables and red lines refer to gas variables. Note that in the density profile, the gas and dust densities are normalised by their own preshock density, which differs in the $D=0.01$ case.}
  \label{fig:accretion_profile}
\end{figure}

We consider the effect of changing the dust to gas ratio. The canonical value in the interstellar medium is $D=0.01$, however values as high as unity have been used in protostellar discs \citep[e.g.][]{dipierro_planet_2015} to account for dust migration to the inner parts of the disc. The velocity, density, and temperature profiles for accretion shocks with $D=0.01$ and $D=1$ are shown together in fig.~\ref{fig:accretion_profile} in the case where the dust size $r_d=10$~$\mu$m. The profiles are similar, with the larger dust-to-gas ratio resulting in a more compressed structure. The main difference is in the dust temperature profile (lower panel). When $D=1$, the dust does not heat above its preshock value of 10~K. This means that if an accretion shock is observed, the dust temperature is a probe of the dust-to-gas ratio. Even though this ratio changes within the shock, it always returns to the preshock value. Hence the peak dust temperature measures the dust-to-gas ratio of material that eventually falls onto the protoplanetary disc.

We have presented an astrophysical application of two-fluid dust-gas shocks by studying the accretion shock above a protoplanetary disc. We have simplified the system by not including any chemical reactions or evaporation of grain mantles. At the temperatures reached ($10^3$~K) there is significant driving of neutral-neutral reactions, and coolants such as H$_2$O and OH could be produced. Cooling by these molecules could change the detailed structure of the shock and/or provide radiative signatures of the shock parameters. Our simple treatment has shown that a detailed analysis of dust-gas shocks could be useful to investigate infalling material onto protoplanetary discs.

\section{CONCLUSION}
We have numerically solved the two-fluid dust-gas equations assuming a steady-state, planar structure. Two distinct shock solutions exist where the gas fluid drags the dust fluid along through a discontinuity (J-type) or smoothly (C-type) until both fluids settle onto post-shock values. These shocks are ideal tests for benchmarking the behaviour of numerical codes seeking to simulate dusty gas with different expressions for the drag or dust-to-gas mass density ratios. Our \textsc{python} code that returns shock solutions for user defined parameters is publicly available on the Python Package Index\footnote{https://pypi.python.org/pypi/DustyShock} and BitBucket\footnote{https://bitbucket.org/AndrewLehmann/dustyshock}.

We used a J-type two-fluid dust-gas shock to study the accretion shock settling material onto a protoplanetary disc. We found that two-fluid effects are most likely to be important for larger grains ($r_d > 1$~$\mu$m). The dust temperature within the shock front was found to be a sensitive probe of the dust-to-gas ratio that eventually falls onto the protoplanetary disc. This work shows that a detailed analysis of two-fluid dust-gas shocks could be a fruitful avenue to investigating the composition of infalling material onto protoplanetary systems.

\section*{Acknowledgments}
The authors gratefully acknowledge discussions with James Tocknell, Shane Vickers and Birendra Pandey. This research was supported by the Australian Research Council through Discovery Project grant DP130104873. AL was supported by an Australian Postgraduate Award.


\bibliographystyle{mnras}
\bibliography{dustyshocks}

\end{document}